\documentclass[
    ,final            
  ]
  {aipproc}

\layoutstyle{6x9}

\begin{document}

\title{Testing Binary Population Synthesis Models with Hot Subdwarfs}

\classification{95.85.Mt; 97.80.Fk}
\keywords {binaries: close -- stars: horizontal-branch -- subdwarfs -- ultraviolet: stars}

\author{Richard A.\ Wade}{
  address={Dept.\ of Astronomy \& Astrophysics, Pennsylvania State
  University , 525 Davey Lab, University Park PA 18602 USA} }

\author{Ravi kumar Kopparapu}{
  address={Institute for Gravitational Physics and Geometry,
Pennsylvania State University, 334 Whitmore Lab,
University Park, PA 16802-6300, USA}
}

\author{Richard  O'Shaughnessy}{
  address={Institute for Gravitational Physics and Geometry,
Pennsylvania State University, 334 Whitmore Lab,
University Park, PA 16802-6300, USA}
}

\begin{abstract}
Models of binary star interactions have been successful in explaining
the origin of field hot subdwarf (sdB) stars in short period systems,
but longer--period systems that formed {\em via} Roche--lobe overflow
(RLOF) mass transfer from the present sdB to its companion have received
less attention.  We map sets of initial binaries into present--day
binaries that include sdBs and main--sequence stars, distinguishing
``observable'' sdBs from ``hidden'' ones. We aim to find out whether (1) the
existing catalogues of sdBs are sufficiently fair samples of all the
kinds of sdB binaries that theory predicts; or instead whether (2) large
predicted hidden populations mandate the construction of new catalogues,
perhaps using wide--field imaging surveys such as 2MASS, SDSS, and Galex.
We also report on a pilot study to identify hidden subdwarfs, using
2MASS and GALEX data.

\end{abstract}

\maketitle


\section{FORMATION CHANNELS FOR HOT SUBDWARF BINARIES}

Hot subdwarf B (sdB) stars are considered to be He--core burning systems
with masses usually near 0.5~M$_\odot$ and thin hydrogen envelopes,
located at the blue end of the horizontal branch. Since they form a
narrow sequence, they are also referred to as Extended Horizontal Branch
(EHB) stars.  Studies by Maxted et al.\ \cite{maxted2001} indicated that
nearly half of the apparently single sdB stars are in short period
binaries with white dwarf (WD) or M dwarf companions. The possible
formation channels include stable, conservative mass transfer through a
Roche lobe overflow (RLOF) or a dynamical--timescale mass transfer
(unstable) leading to a common envelope (CE) evolution. Maxted's
binaries are products of CE evolution. The RLOF channel can generate
luminous main sequence (MS) companions at long periods, with sdB masses
and luminosities lower than typical.  How important is the RLOF channel?

We used the Binary Stellar Evolution (BSE) code of Hurley et al.\
\cite{hurley2002}, mapping sets of initial binaries into present--day
binaries that include sdBs, and distinguishing ``observable'' sdBs from
``hidden'' ones. The common envelope efficiency parameter, $\alpha_{CE}$
was set to 1.5. One million initial binaries were chosen at random using
an initial mass function for $m_2$, uniform distribution of mass ratio
$m_1/m_2 < 1$, and log--uniform distribution of initial separation.  We
define a star as an sdB in the BSE code if it has surface gravity in the
range $5.0 < \log g < 6.6$ (cgs) and effective temperature in the range
$20,000~{\rm K} < T_{\rm eff} < 45,000$~K. Systems that form sdB binaries
with MS companions have initial values of $m_1$ between 0 and 5
M$_\odot$ depending on $m_2$.  Initial orbital periods for these systems
are between 1 and 200 days.

The distribution of $T_{\rm eff}$ for the MS companions versus the
orbital period after sdB formation shows two distinct sub--populations,
reflecting the RLOF channel (long period, 2--300 d) and the CE channel
(short period, 0.3--3 d). Similar work by Han et al.\ \cite{han2003}
shows an absence of systems with A--F type companions for orbital
periods shorter than 10 days. Using the Hurley et al.\ \cite{hurley2002}
code, we do not find such a gap.  Observations of a ``fair sample'' of
sdB binaries should provide some insights.

\section{AN EXAMPLE BINARY WITH A HIDDEN HOT SUBDWARF}

\begin{figure}
  \includegraphics[height=.33\textheight]{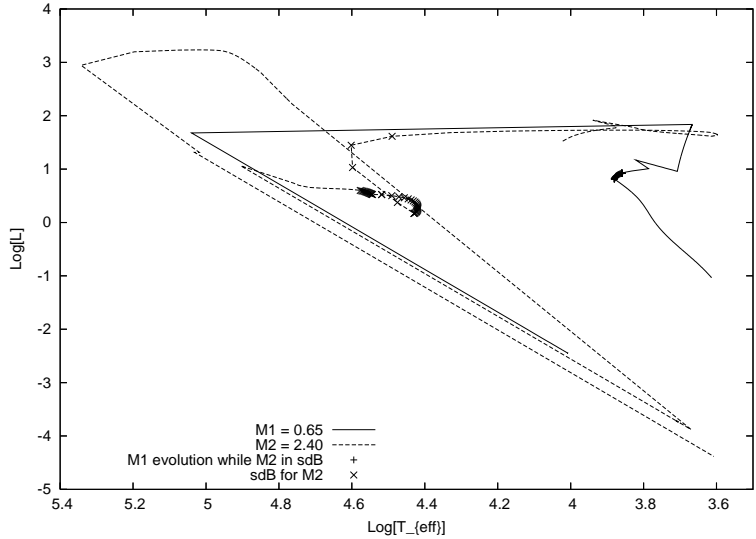}
  \caption{A ($\log T_{\rm eff}, \log L$) diagram illustrating the evolution
  of a binary system that forms an sdB star and an F0 companion.}
\end{figure}

Figure 1 shows the stellar evolutionary tracks for a binary system that
forms an sdB star and a MS companion; initial masses were 0.65 and 2.40
M$_\odot$, the initial orbital period was 2.20 days.  The more massive
star (dashed line) evolves off the MS first and fills its Roche lobe to
start a stable mass transfer onto its low mass MS companion, which
becomes an F0 star. In the process, the first star loses its envelope to
form a He--burning star, which we classify here as an
sdB. Crosses/plusses mark the parts of the tracks of both stars while
the initially more massive star is in the sdB phase. This stage of
sdB+F0 MS binary lasts for $\sim$1.0 Gyr, so such systems may be
relatively numerous.  The sdB stars exhausts He and becomes a CO WD.
The F0 star remains on the MS for another $\sim$1.2 Gyr.  The system
evolves further, after a CE episode, as a pair of low--mass WDs (CO +
He) which coalesce at age 3.3 Gyr, briefly making a He star, then a CO
WD.

Figure 2 presents brightness estimates at different wavelengths of the
sdB star and its F0 companion from Fig.~1. The sdB star has lower than
canonical mass (0.319 M$_\odot$), hence low luminosity and a long
lifetime.  The system has been placed at 1.0 kpc.  Vertical bars
indicate the approximate effective wavelengths and useful magnitude
ranges of GALEX AIS, SDSS, and 2MASS.  During the long sdB+F0 phase, the
companion dominates the sdB in the optical, but not in the far UV.

\begin{figure}
  \includegraphics[height=.3\textheight]{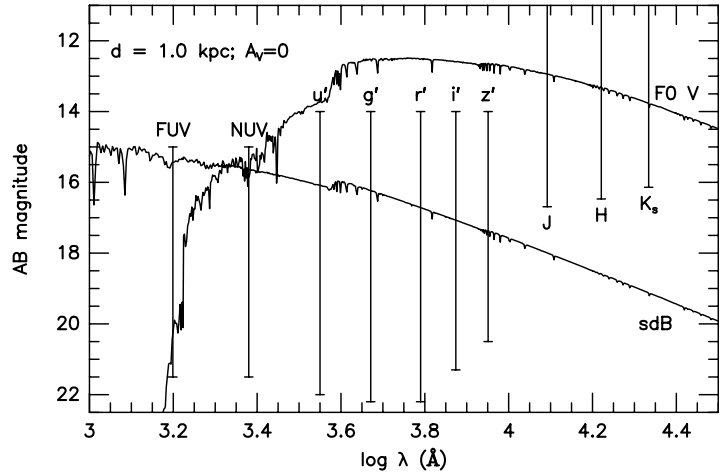} 
\caption{The
  spectral energy distributions of the sdB and F0 stars from Fig.\ 1.
  Vertical bars indicate the approximate effective wavelengths and
  useful magnitude ranges of GALEX AIS, SDSS, and 2MASS.}
\end{figure}

Such systems typically would not be found in present catalogues of sdB
stars.  In the SDSS u'g'r'i'z' bandpasses, the companion star would be
saturated at 1.0 kpc. At d $>$ 1.0 kpc, where SDSS is not saturated,
either one must observe near the Galactic plane (where SDSS and GALEX do
not go and where reddening complicates the analysis) or one must observe
several scale heights above the disk.  A more effective way to search
for such systems is to select F--star candidates based on their near--IR
(2MASS) colors, then search for a far--UV excess (UVX) using GALEX.  A
preliminary search for F+UVX systems near $V\sim12$ suggests that 1 or 2
F stars out of 100 may be hiding a hot subdwarf companion.  We are
observing a few of these at the Hobby--Eberly Telescope, to see whether
the F star's radial velocity is variable. The ultimate goal is to
measure the ratio of sdB binaries formed by the RLOF and CE channels in
a fair sample, and to find the critical mass ratio that separates stable
from unstable mass transfer.


\begin{theacknowledgments}
Supported by NASA NNG05GE11G and NSF PHY 03-26281, PHY 06-00953 
and PHY 06-53462. This work was also supported by the Center
for Gravitational Wave Physics, which is supported by the National
Science Foundation under cooperative agreement PHY 01-14375.
\end{theacknowledgments}


\end{document}